\documentclass[showpacs,showkeys]{revtex4}
\usepackage{graphicx}
\usepackage{dcolumn}
\usepackage{bm}

\begin{document}

\title{ Directed Coulomb explosion effect on proton acceleration
by an intense laser pulse from a double-layer target }

\author{Toshimasa~Morita}
\author{Sergei~V.~Bulanov}
\author{Timur~Zh.~Esirkepov}
\author{James~Koga}
\author{Masaki~Kando}

\affiliation{Advanced Beam Technology Research Division,
Japan Atomic Energy Agency, 8-1-7 Umemidai, Kizugawa,
Kyoto 619-0215, Japan}


\begin{abstract}
We examine ion acceleration by irradiating a hundred TW laser pulse on a
double-layer target.
It is shown analytically and by three-dimensional particle-in-cell simulations
that higher energy protons are obtained by using
material with a high charge-to-mass ratio in the
first layer of a double-layer target,
because a strong Coulomb explosion occurs
in such a material.
As a result, the protons keep accelerating for a longer time.
Using the optimal conditions for the target,
it is shown that high energy and high quality protons can be generated.

\end{abstract}


 
\maketitle

Laser driven charged particle acceleration is one of the important
examples of applying compact laser technology.
This method of fast particle generation
is very attractive, since the acceleration rate is markedly higher and the
facility size can be substantially smaller than that of conventional
accelerators.
Laser driven fast ions are promising in many applications such as
hadron therapy \cite{SBK},
fast ignition of thermonuclear fusion \cite{ROT}, production of
positron emission tomography (PET)
sources \cite{SPN}, conversion of radioactive waste \cite{KWD}, proton
imaging of ultrafast processes in laser plasmas \cite{PrIm}, a laser-driven
heavy ion collider \cite{ESI1}, and a proton dump facility for neutrino
oscillation studies \cite{SVB}.

Although there are many experimental studies of laser acceleration of ions
(see review articles \cite{BOR}),
the achieved proton energy at present is not high enough
for some applications such as hadron therapy
which requires one-two hundred MeV protons.
Higher energy protons can be obtained by using a higher power laser.
However, the laser power enhancement
will result in the cost increase of the accelerator.
Therefore, it is important to study the conditions for generating higher
energy protons with a lower power laser by using some special techniques.
For example, one can obtain the required energy protons by using
radiation pressure dominant acceleration (RPDA) and its modifications
\cite{ESI1,SSB,TAG,ZCW}.
The laser peak intensity is $I_0 \geq 10^{22}$W/cm$^{2}$ in those studies.

\begin{figure}[tbp]
\includegraphics[clip, width=10.0cm, bb=25 64 555 358]{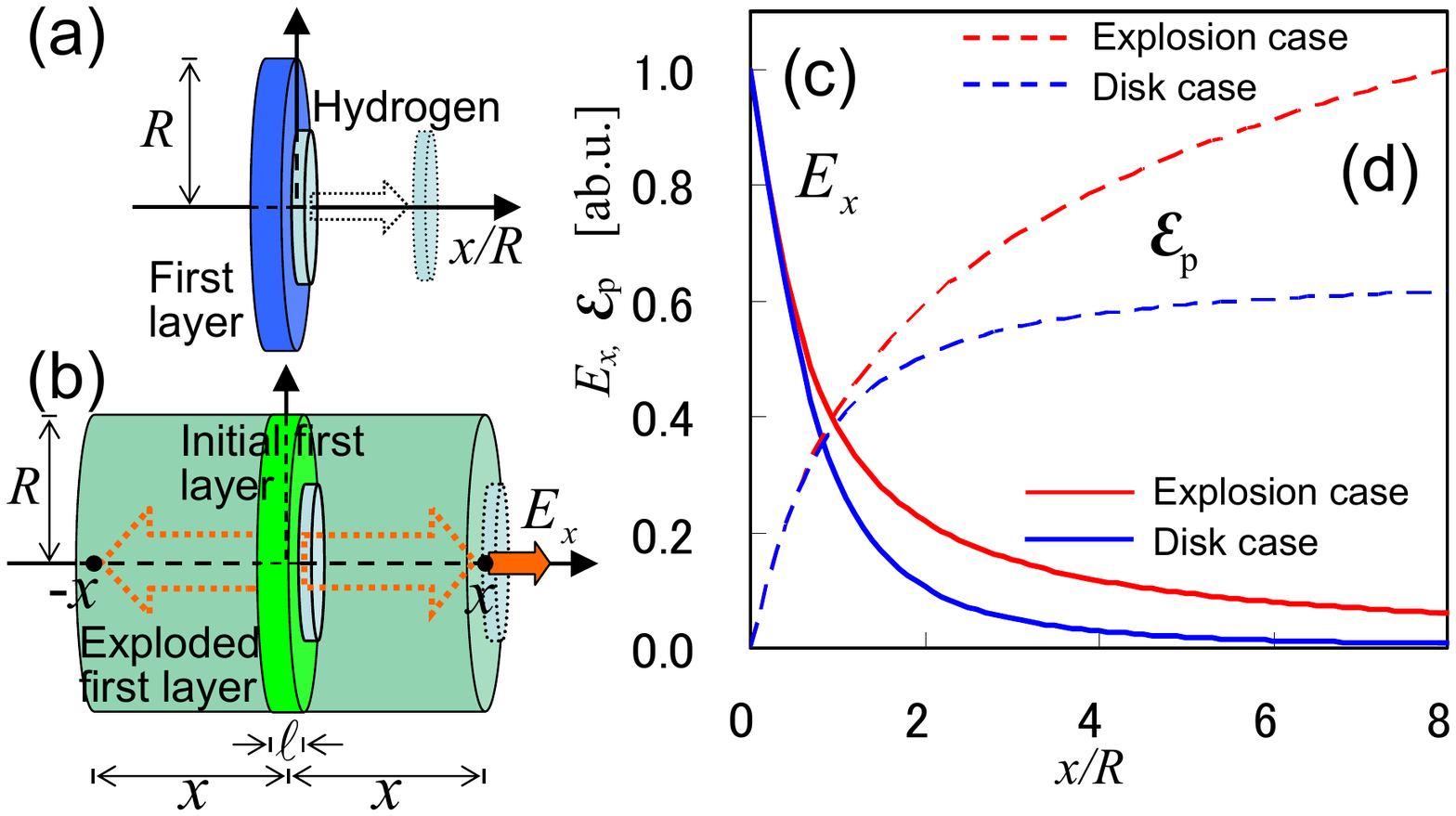}
\caption{\label{fig:fig01}
(a) The first layer and proton bunch shape of the disk case.
(b) The initial shape of the target and
the exploded first layer by the Coulomb explosion.
(c) The electric field component, $E_x$, normalized by $E_0$, of the
disk case (blue solid line) and the Coulomb explosion case (red solid line).
(d) The energy gain of protons, $\mathcal{E}_{p}$,
normalized by the maximum of the Coulomb explosion case,
the disk case (blue dash line) and the Coulomb explosion case (red dash line).
}
\end{figure}

In this Letter,
we show a method to obtain the high energy protons by using a
laser whose intensity is $I_0 \approx 10^{21}$W/cm$^{2}$
and energy is $\mathcal{E}_{las} \leq 20$J,
when the RPDA regime is not reached in full scale.
We use three-dimensional (3D) particle-in-cell (PIC) simulations
in order to investigate how high-energy and high-quality protons
can be generated by a few hundred TW laser.
We study the proton acceleration during the interaction of a laser pulse with
a double-layer target consisting of some high-$Z$ atom layer and a hydrogen
layer (see Fig. \ref{fig:fig01}(a)). 
As suggested in Refs. \cite{SBK,DL},
a quasimonoenergetic ion beam can be obtained using targets of this type.
Our aim is to obtain a high energy ($\mathcal{E} \geq 100$MeV) and
high quality ($\Delta \mathcal{E}/\mathcal{E}\leq 5\%$)
proton beam using a relatively moderate power laser ($P \approx 500$TW).
We show the dependence of the proton energy
on the material of the first layer.
In experiments of laser driven ion acceleration,
the case of using a CH polymer target observed higher
energy protons \cite{SNAV}
than the case of using a metallic target \cite{CLAR}.

Here we estimate the required laser energy to generate protons with energy
$\mathcal{E} \geq 100$MeV.
The electric field of a charged infinite sheet is equal to
$E_{0}=\rho l/2\epsilon_0$,
where $\rho$ is the charge density, $l$ is the sheet thickness
and $\epsilon_0$ is the vacuum permittivity.
The $x$ component of the electric field of the positively charged disk
is equal to
\begin{equation}
E_x(x)=E_{0} \left(1-\frac{x}{\sqrt{x^{2}+R^{2}}} \right),
\label{exx}
\end{equation}
where $R$ is the charged disk radius.
We assume that the $x$ axis is normal to the disk surface
placed at the disk center.
The energy gain of protons in this electric field is
$\mathcal{E}_p=\int_0^\infty q_eE_x(x)dx=q_eE_{0}R$,
where $q_e$ is the electron charge.

Let $N_e$ electrons be extracted from the initially neutral charge disk.
The resulting charge density is given by
$\rho=q_e N_e/\pi R^2 l$,
where $l$ is the disk thickness.
The energy of the proton is given by
$\mathcal{E}_p=q_e^2N_e/2\epsilon_0\pi R$.
We obtain
\begin{equation}
  N_e=\frac{2\epsilon_0\pi R\mathcal{E}_p}{q_e^2}.
\label{nne}
\end{equation}
This expression shows the number of electrons that should be removed
from the disk to generate
an electrostatic field which accelerates protons up to the
energy of $\mathcal{E}_p$.
Next, we estimate the energy necessary to remove $N_e$ electrons
from the initially neutral charge disk
provided that the remaining ions do not move significantly.
Let $N$ electrons be already removed from the disk.
The charge density of this disk is,
$\rho(N)=q_e N/\pi R^2 l$, and
$E_0(N)=\rho(N)l/2\epsilon_0=q_eN/2\epsilon_0\pi R^2$.
The energy necessary to remove one electron from the disk is
$\mathcal{E}_e(N)=q_{e}E_{0}(N)R=q_e^{2}N/2\epsilon_0\pi R$.
If $N_e$ electrons are removed by the laser pulse,
the required energy of the laser pulse is
\begin{equation}
  \mathcal{E}_{las}=\int_0^{N_e} \mathcal{E}_e(N)dN=
  \frac{\epsilon_0\pi R\mathcal{E}_p^2}{q_e^2}.
\label{elas}
\end{equation}
This expression shows
the minimum laser pulse energy needed to generate a proton with the enegy of
$\mathcal{E}_p$ in the electrostatic field of the disk.
When we assume the diameter of disk to be $10\mu$m,
the necessary laser energy to generate the proton of
$\mathcal{E}_p$=200MeV is calculated to be $\mathcal{E}_{las} \approx$6J.
This estimation assumes that the energy of the laser is
efficiently spent on extracting electrons while ions remain at rest.

In the model described above,
the protons are accelerated in the electric field given by Eq. (\ref{exx}).
This electric field decreases rapidly with the distance
from the disk surface as shown in Fig. \ref{fig:fig01}(c) (blue solid line). 
To enhance the ion acceleration,
we must produce the condition where the ions are being accelerated for a longer
time by the electric field which more slowly decreases with distance
as seen by the protons.
This can be achieved using a moving electrostatic potential,
associated with the motion of the first layer of the double-layer target.
With a sufficiently high power laser the first layer can be pushed forward
in the RPDA regime \cite{ESI1,SSB}.
However, as suggested in \cite{SSB},
a Coulomb explosion of the first layer can produce a moving electrostatic
potential, by which the second layer is accelerated.

Here we estimate the effect of this Coulomb explosion of the first layer.
To simplify the discussion,
we assumed that the disk target expands in the shape of cylinder.
It is assumed that the target thickness, $l$, of the initial shape
expanded to $+x$ and $-x$ and the target radius, $R$, expanded to $r(x)R$,
where $r(x) \ge 1$, 
and the distance of a proton from the expanding target surface is $x'$.
The electric field $E_x$ at the proton position $x_p=x+x'$ is 
\begin{equation}
E_x(x_p) = \frac{E_0}{r^2(x)}
  \left\{ 1 + \sqrt{ \left(\frac{x'}{2x}\right)^2
           + \left(\frac{r(x)R}{2x}\right)^2 } 
         - \sqrt{ \left(1+\frac{x'}{2x}\right)^2
         + \left(\frac{r(x)R}{2x}\right)^2}
  \right\},
\label{exdx}
\end{equation}
where $E_0$ is the value for the initial disk.
When the $r(x)$ is a linear function of $x$ of the form $r(x)=ax+1$,
$a$ is a real number and $a \ge 0$,
and the proton is on the expanding target surface,
this electric field is
$E_x(x)=E_0\{1+aR/2+R/2x-\sqrt{1+(aR/2)^2+aR^2/2x+(R/2x)^2}\}/(ax+1)^2$.
Here $E_x(x)$ and $\mathcal{E}_{p}(x)$ increase with the decrease of $a$.
In the limit of $a=0$, $r(x)=1$ (Fig. \ref{fig:fig01}(b)),
this electric field is
$ E_x(x)= E_0( 1+R/2x-\sqrt{1+(R/2x)^2})$.
This electric field $E_x(x)$ is shown in
Fig. \ref{fig:fig01}(c) (red solid line).
The electric field of this case is higher than the disk case at all positions.
The energy gain of protons is
$\mathcal{E}_{p}(x)= q_e E_0 \int_{l/2}^{x} E_x(\tilde{x})d\tilde{x}$.
The proton energy of the Coulomb explosion case is 1.6 times higher than the
disk case until the position $x/R$=8 (see Fig. \ref{fig:fig01}(d)).
In a spherical expansion, on the other hand,
the proton energy is the same in the expanding target and non-expanding one.

The acceleration rate is higher when
the front of the expanded first layer has the velocity close
to the accelerating protons.
That is, we should make a strong Coulomb explosion of the first layer.
The higher the $q_i/m_i$ of the ions, the greater their energy.
A stronger Coulomb explosion occurs in the high $q_i/m_i$ material.
Therefore, we can obtain higher energy protons by using
the high $q_i/m_i$ material in the first layer.
This consideration is corroborated by the simulations described below.

The simulations were performed with a 3D massively parallel
electromagnetic code, based on the PIC method \cite{CBL}.
The dimensional quantities are given in terms of
the laser wavelength
$\lambda = 0.8\mu $m; the spatial coordinates are normalized by $\lambda $
and the time is measured in terms of the laser period, $2\pi/\omega$.
We use an idealized model, in which a Gaussian $p$-polarized laser pulse is
incident on a double-layer target.
The laser pulse with the dimensionless amplitude
$a=q_eE_{0}/m_{e}\omega c=50$, which corresponds to the laser peak intensity of
$5\times 10^{21}$W/cm$^{2}$, is $10\lambda $ long in the propagation direction
and is focused to a spot with size $4\lambda $ (FWHM),
which corresponds to the laser power of $620$TW and laser energy of $18$J.
Both layers of the double-layer target are shaped as disk.
The first layer has the diameter of $8\lambda $ and thickness of
$0.5\lambda $.
The second, hydrogen,
layer is narrower and thinner; its diameter is $4\lambda $
and thickness is $0.03\lambda $.
Although, the first layer material is varied,
it is assumed that it is comprised of ions with the charge state of $Z_{i}=+6$
and the number of ions is the same in all cases. 
The electron density inside the first layer is $n_{e}=3\times 10^{22}$cm$^{-3}$
and inside the hydrogen layer is $n_{e}=9\times 10^{20}$cm$^{-3}$.
The total number of quasi-particles is $8\times 10^{7}$.

\begin{figure}[tbp]
\includegraphics[clip,width=10.0cm, bb=25 60 545 415]{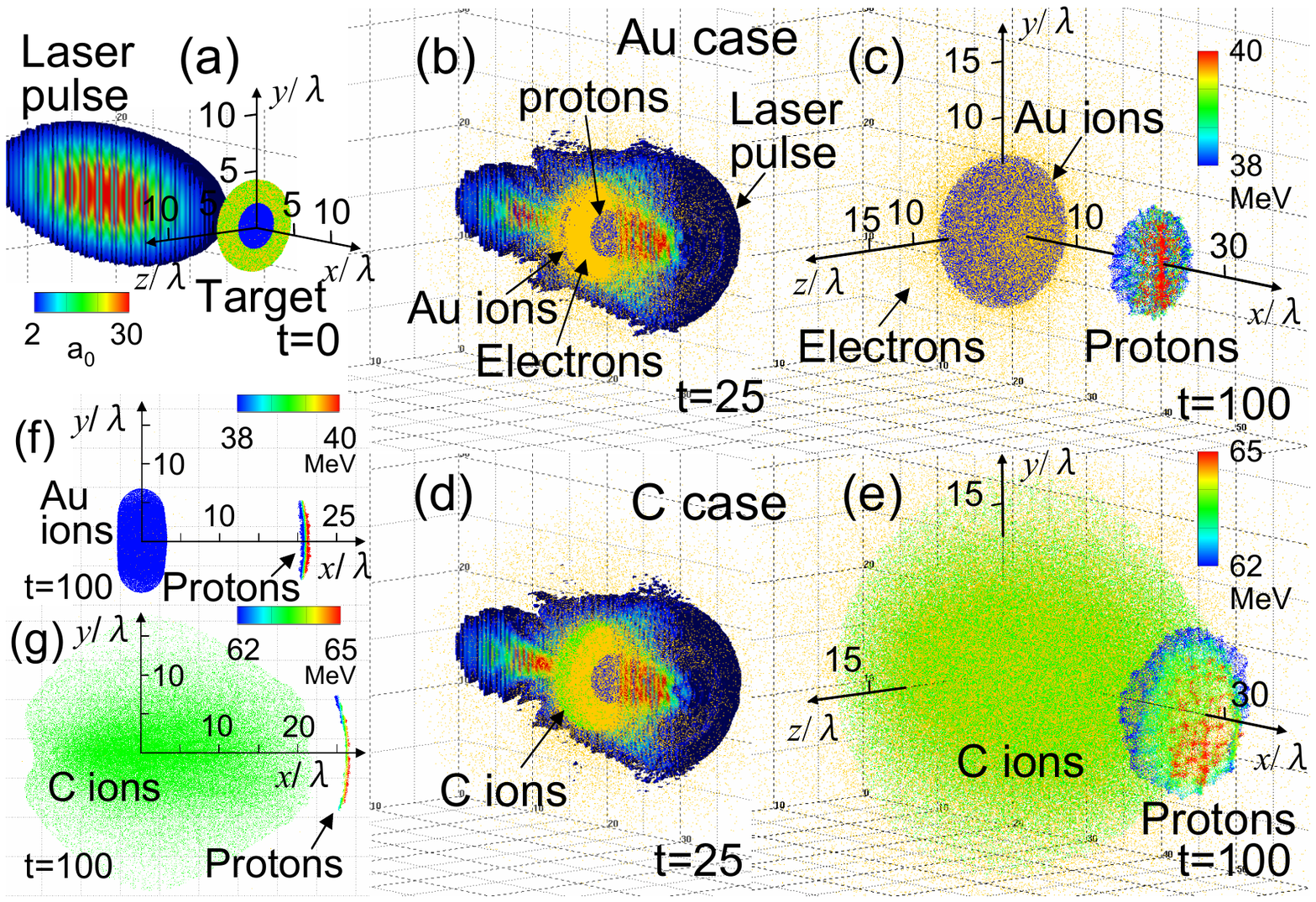}
\caption{\label{fig:fig02}
(a)-(e) Particle distribution and electric field magnitude
(isosurface for value $a=2$); half of the box of the electric field is removed
to reveal the internal structure.
(a) Initial shape of the target and the laser pulse,
(b),(d) the interaction of the target and laser pulse,
and (c),(e) the first layer shape and the accelerated protons (color scale).
(b),(c) The gold case and the (d),(e) carbon case.
(f) Distribution of gold ions and protons of the gold case,
and (g) the carbon case; a 2D projection is shown looking along the $z$ axis.
}
\end{figure}

\begin{figure}[tbp]
\includegraphics[clip, width=10.0cm, bb=26 59 396 365]{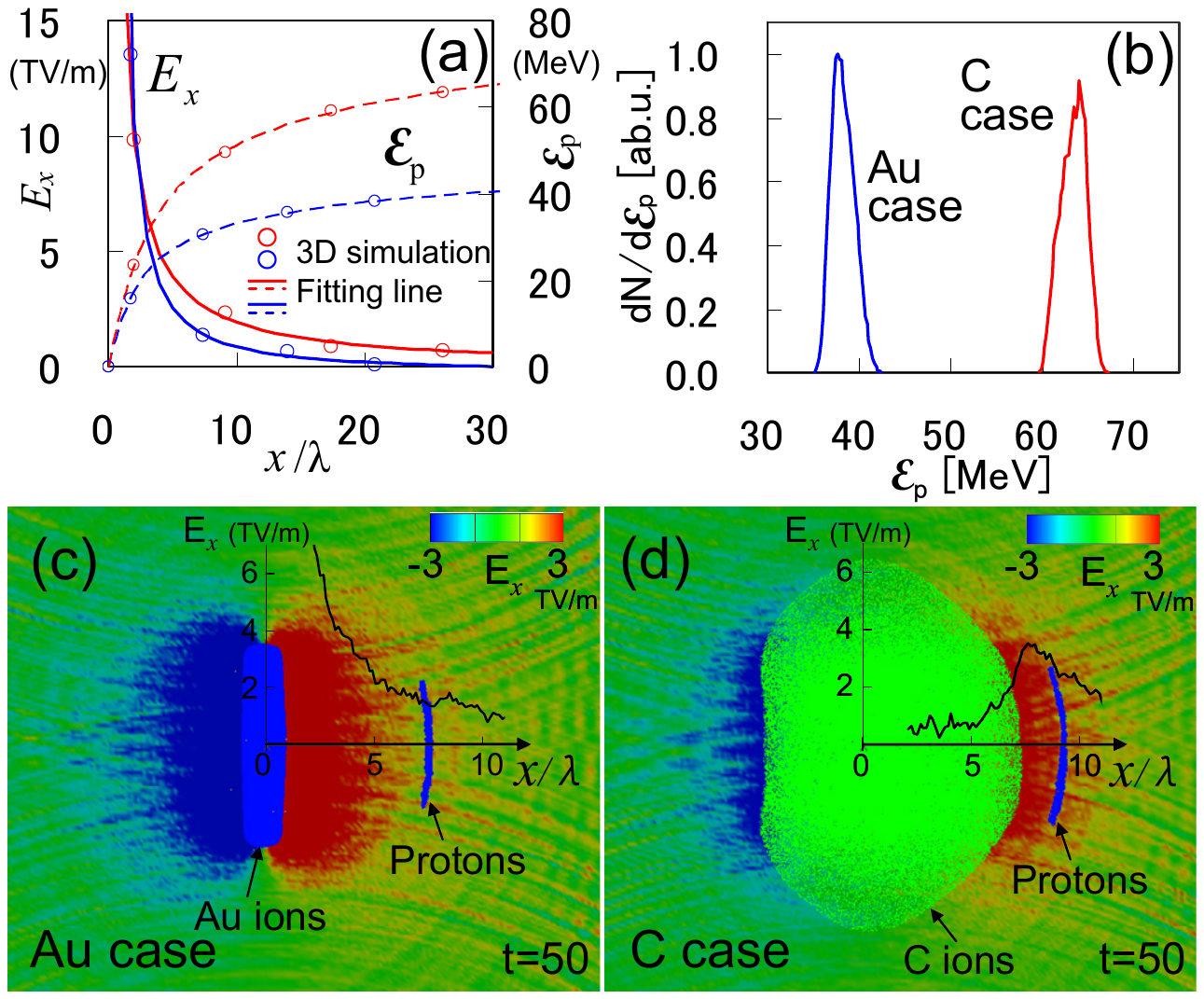}
\caption{\label{fig:fig03}
(a) The accelerating electric field, $E_x$, that the protons receive and
the energy gain of protons $\mathcal{E}_{p}$.
(b) The proton energy spectrum of the gold case and
the carbon case at t=100, normalized by the gold case maximum peak.
(c)(d) The spatial distribution of particles and
the electric field component $E_x$ (color scale) and
the electric field component $E_x$ on the $x$ axis at $t=50$.
}
\end{figure}

The laser pulse is normally or obliquely incident on the target.
The oblique incidence of the laser pulse is realized by rotating the target
about the $Z$ axis.
In the normally incident cases,
the number of grid cells is equal to $3300\times1024\times 1024$
along the $X$, $Y$, and $Z$ axes.
Correspondingly, the simulation box size is
$120\lambda \times 36.5\lambda \times 36.5\lambda$.
In the obliquely incident cases,
the dimension of $L_Y$ is set to be two times larger
than the normally incident cases \cite{MEBKY2}.
The boundary conditions for the particles and for the fields are
periodic in the transverse ($Y$,$Z$) directions and absorbing at the
boundaries of the computation box along the $X$ axis.
In figures \ref{fig:fig02},\ref{fig:fig03},\ref{fig:fig05},
the $x$ axis denotes the direction perpendicular to the target surface and
the $y$ axis is parallel to the target surface,
while the direction of $z$ axis is the same as the direction of the $Z$ axis.
The origin in the $xyz-$coordinates is located at the center of the rear surface
of the initial first layer.

We show two results, one is a case using gold for the first layer
and the other case uses carbon.
In order to examine the dependence of the proton energy $\mathcal{E}_{p}$
versus the first layer material,
we performed simulations with the normal incidence of laser pulse.
Figures \ref{fig:fig02}(b),(c) show the case of gold for the first layer
and Figs. \ref{fig:fig02}(d),(e) show the case of carbon.
We see that the carbon ions are distributed over a wider area than
in the gold case by the Coulomb explosion.
Figures \ref{fig:fig02}(f),(g)
show a 2D projection onto the $(x,y)$ plane.
Here we present the region of $z=-0.7\sim 0.7\lambda$
to see the ion density around the $z=0$ plane.

In the carbon case, the center of the exploded first layer moves in the laser
propagation direction (Fig. \ref{fig:fig02}(g)).
This velocity is about $1/10$ of that of the protons.
We show the estimation of this moving first layer effect on the proton energy.
We assume that the first layer velocity is $V$ and
the second layer ion, proton, velocity is $v$.
That is, the accelerating electric field moves at speed $V$,
it is denoted  $E_x(x-Vt)$.
For $mV^2 \ll \mathcal{E}_{p0}$, we obtain the proton energy
\begin{equation}
\mathcal{E}_p=\mathcal{E}_{p0}+V\sqrt{2m\mathcal{E}_{p0}},
\label{ekv}
\end{equation}
where $\mathcal{E}_{p0}$ is the proton energy
in the case of a non-moving first layer,
and $m$ is the proton mass.
For the proton velocity in the non-moving first layer $v_0 \ll c$ we obtain
$\mathcal{E}_p=\mathcal{E}_{p0}(1+2V/v_0)$.
$V/v \approx 0.1$ in our simulation,
$v/v_0 \approx 1.1$,
so we obtain $20\%$ higher energy protons by this moving first layer.
\begin{figure}[tbp]
\includegraphics[clip,width=6.5cm, bb=52 80 548 527]{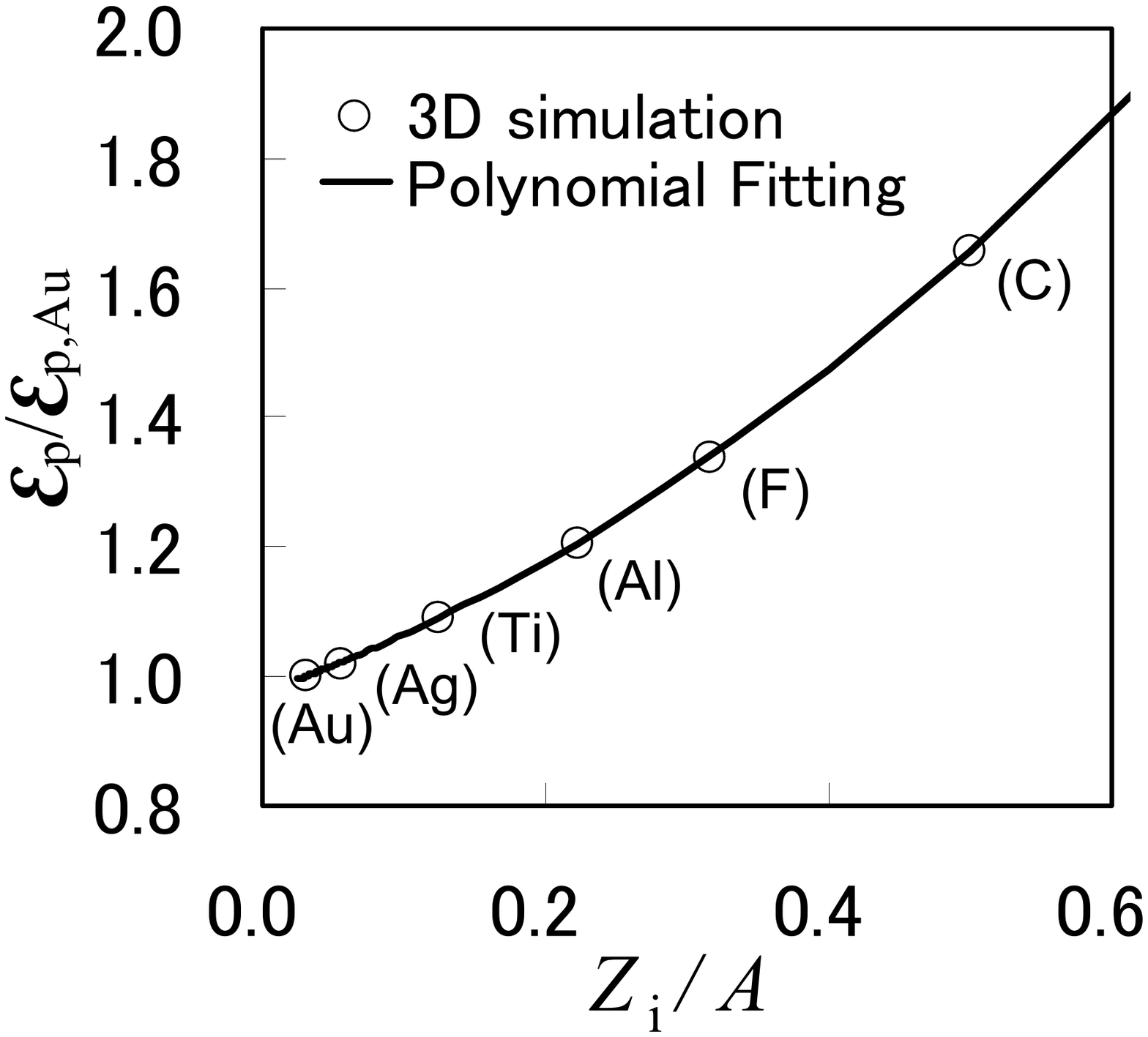}
\caption{\label{fig:fig04}
Proton energy normalized by the energy of the gold case vs
the first layer material.
}
\end{figure}

Figure \ref{fig:fig03}(a) shows the simulation results of the accelerating
electric field that the protons see in the gold and carbon case,
and the energy gain of the protons in each case.
The accelerating electric field of the carbon case is higher than the gold
case and the energy gain of the protons in the carbon case is higher too.
It is similar to the theoretical lines (Fig. \ref{fig:fig01}(c)(d)).
Figure \ref{fig:fig03}(b) shows the energy spectrum of the generated protons
for the gold case and the carbon case.
The average proton energy, $\mathcal{E}_\mathrm{ave}$, is 38MeV in the
gold case and 63MeV in the carbon case.
The proton energy in the carbon case is 1.7 times higher than the
carbon case.
The proton energy spreads are 8.3$\%$ in the gold case and 5.6$\%$ in the
carbon case.
Figures \ref{fig:fig03}(c),(d) show the particle distribution and the $x$
component of the electric field, $E_x$, which accelerates the protons,
around the target at $t=$50.
The black solid lines show the $E_x$ on the $x$ axis.
The value of $E_x$ in the carbon case at the proton's position is $2.4$TV/m
and the gold case is $1.4$TV/m.
That is, the protons in the carbon case undergo a stronger acceleration.

Figure \ref{fig:fig04} shows the average proton energy for some different
materials of the first layer.
The horizontal axis is for $Z_i/A$,
where $Z_i$ is the charge state of the ion and $A$ is the atomic mass number. 
The vertical axis is for the proton energy which is normalized by the energy
in the gold case.
The higher energy protons can be obtained by using a larger ratio of $Z_i/A$.
That is, higher energy protons are obtained by using a
high $q_i/m_i$ material in the first layer of a double-layer target.

\begin{figure}[tbp]
\includegraphics[clip,width=11.0cm, bb=28 67 553 403]{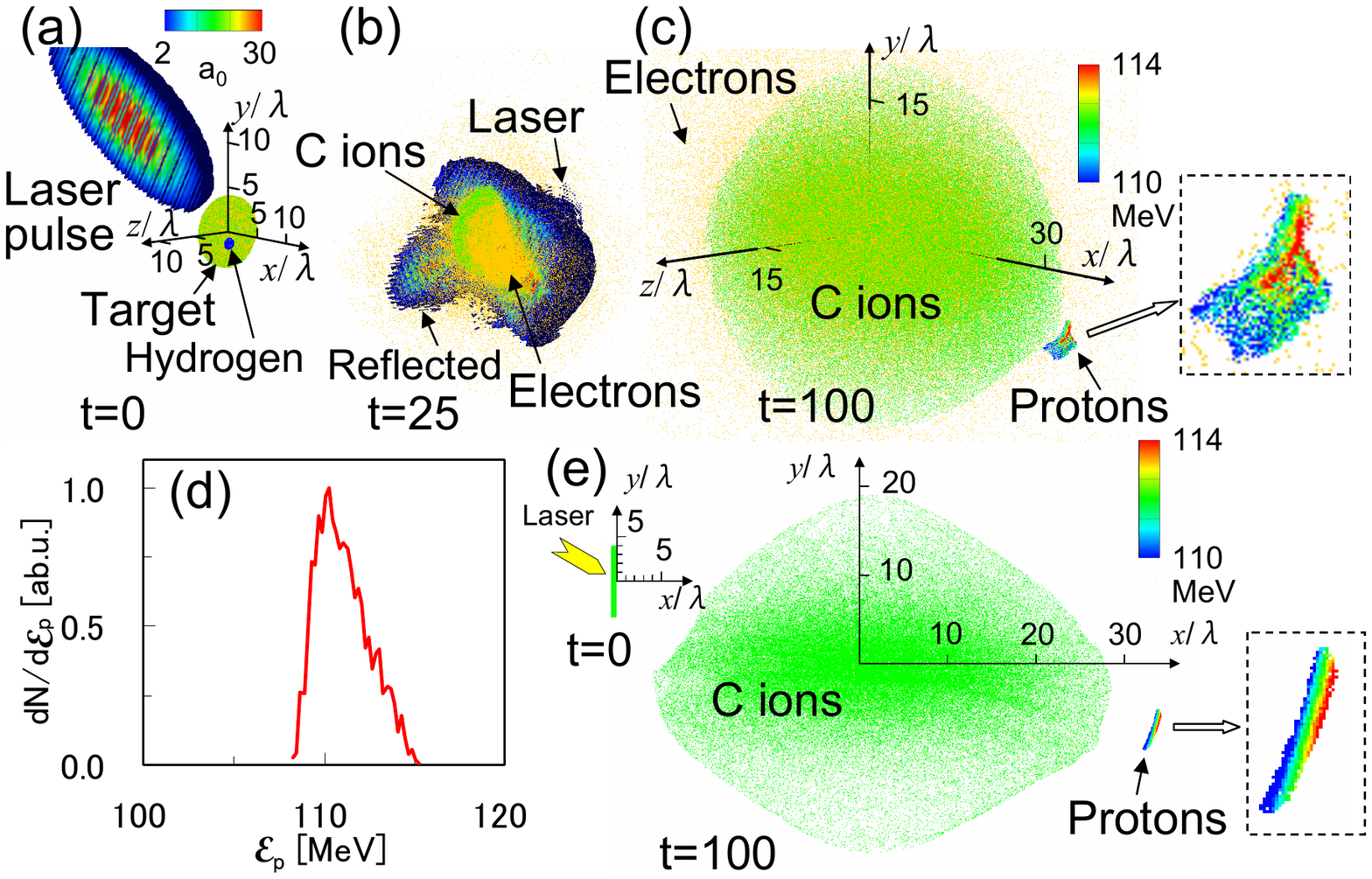}
\caption{\label{fig:fig05}
(a) Initial shape of the target and the laser pulse with oblique incidence,
(b) the interaction of the target and laser pulse, and
(c) the exploding first layer and the accelerated protons (color scale).
(d) The proton energy spectrum normalized by its maximum at t=100.
(e) Distribution of carbon ions and protons (color scale),
a 2D projection is shown looking along the $z$ axis.
}
\end{figure}

In Fig. \ref{fig:fig05} we show the case of an oblique incidence of the laser
pulse onto the carbon target, showing the highest rate of proton acceleration
in the above consideration.
The laser pulse incidence angle is $30^{\mathrm{o}}$ \cite{MEBKY}.
The small hydrogen layer with the diameter of $4/3\lambda$
is shifted below the target center \cite{MEBKY2} (see Fig.\ref{fig:fig05}(a)).
The other parameters are the same as described above.
Figures \ref{fig:fig05}(b),(c) show
the electric field magnitude and distribution of ions.
Half of the box is removed to reveal the internal structure of the electric
field.
We see that the carbon ions are distributed over a wider area
due to the Coulomb explosion.
Figure \ref{fig:fig05}(d) shows
the energy spectrum of the generated protons at $t=100$.
We obtain protons with an average energy, $\mathcal{E}_\mathrm{ave}$,
of $111$MeV, an energy spread, $\Delta\mathcal{E}/\mathcal{E}_\mathrm{ave}$,
of 2.8$\%$, and the number of generated protons of $2\times10^7$.
Figure \ref{fig:fig05}(e) is
the 2D projection onto the $(x,y)$ plane.

In conclusion,
proton acceleration driven by a laser pulse irradiating a double-layer target,
consisting of some high-$Z$ atom layer and a hydrogen layer,
is investigated with the help of 3D PIC simulations.
We find  that higher energy protons are obtained by using a
material with a high charge-to-mass ratio
in the first layer of a double-layer target.
As seen in our simulations,
due to the strong Coulomb explosion occurring in such a material,
the protons keep accelerating for a longer time.
We show that three times higher energy protons are obtained,
even using the same laser pulse,
by using the optimum material for the first layer
and selecting the optimum incidence angle of the laser pulse.

The computation as performed using the PRIMERGY BX900 supercomputer
at JAEA Tokai.

\end{document}